\documentclass[a4paper,fleqn,usenatbib,useAMS]{mnras}

\usepackage{graphicx}	
\usepackage{amsmath}	
\usepackage{amssymb}	
\usepackage{multicol}        
\usepackage{bm}		
\usepackage{pdflscape}	

\def\dr{{\rm d}}

\def\des{\dr {E}_{\rm sh}}
\def\der{\dr {E}_{\rm rlx}}

\def\ggmc{\gamma_{\rm GMC}}
\def\Mi{M_{\rm i}}

\def\sigrel{\sigma_{\rm rel}}
\def\Siggmc{\Sigma_{\rm GMC}}

\def\rhogmc{\rho_{\rm ISM}}

\def\reff{r_{\rm eff}}
\def\rh{r_{\rm h}}

\def\rhoh{\rho_{\rm h}}

\def\rhohi{\rho_{\rm h,i}}

\def\tdis{\tau_{\rm dis}}

\def\trh{\tau_{\rm rh}}

\def\trs{\kappa}

\def\tshock{\tau_{\rm sh}}

\def\E{\mathcal{E}}

\def\msun{{\rm M}_\odot}

\def\gyr{{\rm Gyr}}
\def\myr{{\rm Myr}}

\def\kms{{\rm{km}\,{\rm s}^{-1}}}
\def\pc{{\rm pc}}

\usepackage[T1]{fontenc}
\usepackage{ae,aecompl}

\usepackage{txfonts}

\title[Collisional evolution of cluster with shocks]
{If it does not kill them, it makes them stronger: collisional evolution of star clusters with tidal shocks}

\author[Mark Gieles and Florent Renaud]
  {Mark Gieles and Florent Renaud\\
Department of Physics, University of Surrey, Guildford, GU2 7XH, UK.\\
}

\date{Accepted 2016 August 12. Received 2016 August 12; in original form 2016 May 19}

\pubyear{2016}

\begin{document}
\label{firstpage}
\pagerange{\pageref{firstpage}--\pageref{lastpage}}
\maketitle

\begin{abstract}
The radii of young ($\lesssim 100\,\myr$) star clusters correlate only
weakly with their masses. This shallow relation has been used to argue
that impulsive tidal perturbations, or `shocks', by passing giant
molecular clouds (GMCs) preferentially disrupt low-mass clusters. We
show that this mass-radius relation is in fact the result of the
combined effect of two-body relaxation and repeated tidal shocks.
Clusters in a broad range of environments including those like
  the solar neighbourhood evolve towards a typical radius of a few
parsecs, as observed, independent of the initial radius.  This
equilibrium mass-radius relation is the result of a competition
between expansion by relaxation and shrinking due to shocks.
Interactions with GMCs are more disruptive for low-mass clusters,
which helps to evolve the globular cluster mass function
(GCMF). However, the properties of the interstellar medium in
high-redshift galaxies required to establish a universal GCMF shape
are more extreme than previously derived, challenging the idea that
all GCs formed with the same power-law mass function.
\end{abstract}




\section{Introduction}
The open clusters in the Milky Way
\citep[e.g][]{2005A&A...438.1163K,2006AJ....131.1559V} and young
clusters in external spirals \citep[e.g.][]{1999AJ....118..752Z,
  2007A&A...469..925S} have radii of a few pc, almost independent of
cluster mass $M$. \citet{2004A&A...416..537L} finds for clusters with
ages $\lesssim 100\,\myr$ and $10^3\lesssim M/\msun\lesssim10^5$ that
the average effective radius $\reff$, defined as the radius containing
half of the light in projection, scales as $\reff\simeq
2.8\,\pc\,(M/10^4\,\msun)^{0.1}$. This is strikingly different from
the mass-radius relation (MRR) of star forming clumps, from which star
clusters presumably form and for which the radius depends strongly on
$M$ \citep{1981MNRAS.194..809L}. For example,
\citet{2014MNRAS.443.1555U} find that the radius of star forming
clumps scales with mass as $3.8\,\pc\,(M/10^4\,\msun)^{0.6}$. It is
not clear whether this difference in MRR of molecular clumps and star
clusters originates from the star formation process that alters the
relation, or whether it results from subsequent evolutionary effects.

The near constant radius of star clusters has important consequences
for their survivability. Interactions with giant molecular clouds
(GMCs) disrupt star clusters \citep{1958ApJ...127...17S}, and this
mechanism has been invoked as an explanation for the dearth of old
open clusters in the Milky Way disc \citep{1985IAUS..113..449W,
  1987MNRAS.224..193T}. \citet{1958ApJ...127...17S} shows that the
corresponding disruption time-scale is proportional to the cluster
density.  \citet{2009ApJ...704..453F} argue that clusters form with
similar densities, such that GMC encounters disrupt clusters
independently of their masses.  However, the weak dependence of
$\reff$ on $M$ of young clusters implies that low-mass clusters are
less dense, and therefore more vulnerable to tidal shocks
\citep[][hereafter G06]{2006MNRAS.371..793G}.

Clusters form in regions with high gas densities and after formation
they drift away from these regions, and cloud interactions are
therefore more important in the early evolution than estimated
  from their current environment
\citep{2010ApJ...712..604E}. \citet{2010ApJ...712L.184E} uses this,
and the observed MRR, to suggest that young globular clusters in the
early Universe with masses of up to $10^5\,\msun$ were more vulnerable
to disruption by gas clouds  than more massive GCs.  Elmegreen
proposes that this early disruption mechanism can evolve a $-2$
power-law cluster mass distribution, as is observed for young massive
clusters \citep*[YMCs;][]{1999ApJ...527L..81Z, 2010ARA&A..48..431P},
into a universally peaked globular cluster mass function 
\citep[GCMF; see also][]{2015MNRAS.454.1658K}.

However, tidal interactions affect not only the clusters' masses, but
also their radii, such that the MRR evolves.  The MRR is also affected
by internal two-body relaxation, which causes clusters to expand until
the galactic tidal field stops the expansion. Here we study cluster
evolution as the result of both tidal shocks and two-body relaxation
by combining prescriptions for the change in the total cluster energy
due to both processes.  The total energy of a self-gravitating stellar
system in virial equilibrium depends on $M$ and the half-mass radius
$\rh$ as $E = -\alpha GM^2/\rh$. Here $G$ is the gravitational
constant and $\alpha$ is a form-factor that depends on the density
profile of the cluster.  Both tidal shocks and two-body relaxation
affect the density profile because of redistribution of energy
\citep*[e.g.][ respectively]{1987degc.book.....S,
  1999ApJ...522..935G}, but for a wide range of cluster models
$\alpha=0.2$ to within 20\% \citep[e.g.][]{1987degc.book.....S}, hence
we fix $\alpha=0.2$ from hereon.  We describe the evolution of the
cluster in terms of $M$ and the average density within $\rh$, $\rhoh$,
so we express $E$ in these quantities: $E \propto
-GM^{5/3}\rhoh^{1/3}$.  The fractional change in $E$ then relates to
variations in $M$ and $\rhoh$ as
\begin{equation}
\frac{\dr{E}}{E} = \frac{5}{3}\frac{\dr{M}}{M} + \frac{1}{3}\frac{\dr\rhoh}{\rhoh}.
\label{eq:de}
\end{equation}
We solve for the evolution of all three variable in two steps: (1)
establish how the evolution of $M$ depends on the evolution of $E$
(independent of time) to find an expression for $\rhoh(E)$; (2) find
the evolution of $E$ on the appropriate time-scales.

In Sections~\ref{sec:shocks} and \ref{sec:relax} we derive the
relations for tidal shocks and relaxation, respectively. In
Section~\ref{sec:both} we combine the two effects and derive an
equilibrium MRR and the evolution of all parameter in time.  Our
conclusions are presented in Section~\ref{sec:conclusions}.

\section{Tidal shocks}
\label{sec:shocks}

\subsection{Density evolution}
\label{ssec:rhoh_E_shocks}
Here we derive the response of a self-gravitating system to a single
tidal perturbation to relate $E$ and $M$. We assume that the duration
of the tidal perturbation is much shorter than the crossing time of
stars in the cluster (i.e. the impulsive regime) such that the effect
of adiabatic damping \citep{1987degc.book.....S,1994AJ....108.1398W,
  1995ApJ...438..702K} can be ignored and the term `shock'
applies. This assumption is justified because typical relative
velocities during encounters with GMCs are much higher ($\gtrsim
10\,\kms$) than the velocities of stars in the {\it outer} parts of
clusters ($\lesssim 1\,\kms$).

The energy gain due to a shock can be expressed in the properties of
the cluster, the GMC and the encounter \citep{1958ApJ...127...17S,
  BT1987}. These analytic results show good agreement with results for
numerical experiments (\citealt*{1999ApJ...513..626G}, G06). G06 also
provide an expression for the mass loss resulting from a single
encounter.  However, neither the theory, nor the numerical work,
provide a description of what the energy of the remaining bound stars
is, which is needed to understand the subsequent response of the
cluster density (see equation~\ref{eq:de}).

\begin{figure}
\includegraphics[width=8cm]{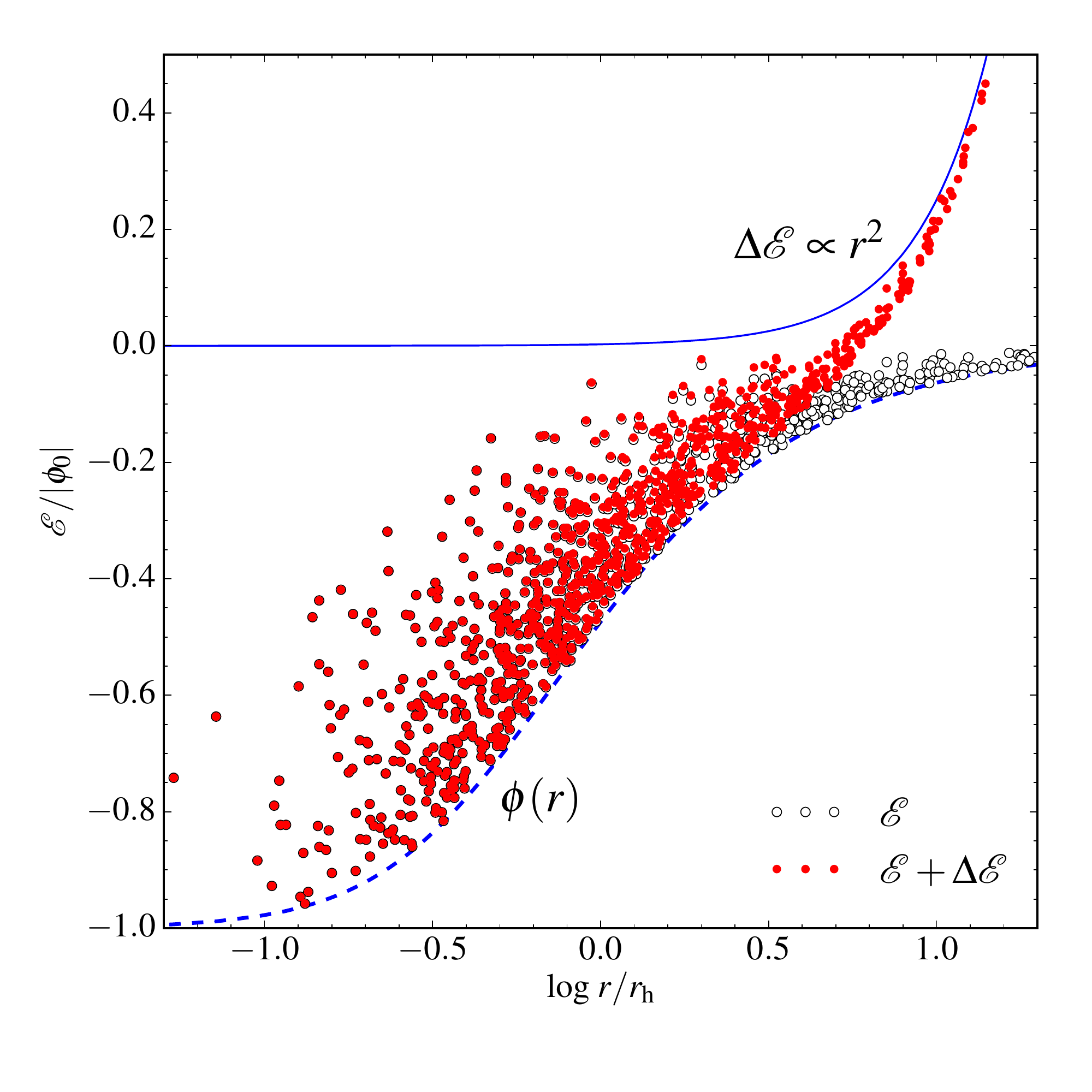}
\caption{Specific energy ($\E$) of $10^3$ stars, normalized to the
  central potential $\phi_0$, as a function of distance to the cluster
  centre in terms of $\rh$, for an isochrone model
  \citep[][open circles]{1959AnAp...22..126H, 1960AnAp...23..474H}. The
  dashed line indicates the specific potential, $\phi$. The filled
  (red) circles represent the energy after a tidal shock, i.e. an
  additional energy $\Delta\E \propto r^2$. The stars that are unbound
  after the perturbation, had an energy close to $0$ before the
  encounter. As a result, the average (specific) energy of remaining
  stars is lower after the shock, and the cluster shrinks.}
\label{fig:schematic_energy}
\end{figure}

To proceed, we introduce a parameter $f$ to relate $E$ and $M$ 
\begin{equation}
\frac{\dr M}{M} = f\frac{\des}{E}.
\label{eq:dmde}
\end{equation}
We assume that all stars are bound to the cluster before a shock is
applied, such that mass loss always results in an increase of $E$
(i.e. $f>0$).  Substituting equation~(\ref{eq:dmde}) in
equation~(\ref{eq:de}) we find an expression for the relation between
$\rhoh$ and $E$

\begin{equation}
\frac{\dr\rhoh}{\rhoh} = \left(3-5f\right) \frac{\des}{E}.
\label{eq:drde_sh}
\end{equation}
Note that combined with equation~(\ref{eq:dmde}) it is straightforward
to express the evolution of $\rhoh$ in terms of $M$.  For $f=3/5$ the
density remains constant and for $f>3/5$ the cluster density goes
up. Similarly, for $f=1/2$ the cluster evolves with a constant $\rh$.

To find an estimate for $f$\footnote{G06 introduced a parameter $f$
that relates energy gain $\Delta E$ to mass loss $\Delta M$ as the
  result of a single encounter between a cluster and a GMC as $\Delta
  M/M_0 = f\Delta E/E_0$, where $M_0$ and $E_0$ are the mass and
  energy of the cluster before the encounter, and they find
  $f\simeq0.2$.  This results applies to the energy gain of all the
  stars, including the unbound stars. Stars escape with positive
  energies, so this result does not give us the required $\Delta E =
  E_1 - E_0$, where $E_1$ is the energy of the remaining bound
  stars.}, we consider the effect of an individual tidal shock.  If a
tidal force works on a cluster for some time, the velocity increase of
a star, $\Delta v$, is proportional to its distance from the cluster
centre $r$. The increase in the specific energy of the stars $\E$ is
then $ \Delta\E \propto r^2$ \citep{1958ApJ...127...17S}, where the
constant of proportionality depends on the strength of the shock. Note
that we ignore the cross term $v\Delta v$, which is small compared to
$(\Delta v)^2$ for escapers, and we refer to
\citet{1999ApJ...513..626G} for a discussion.

\begin{figure}
\includegraphics[width=8cm]{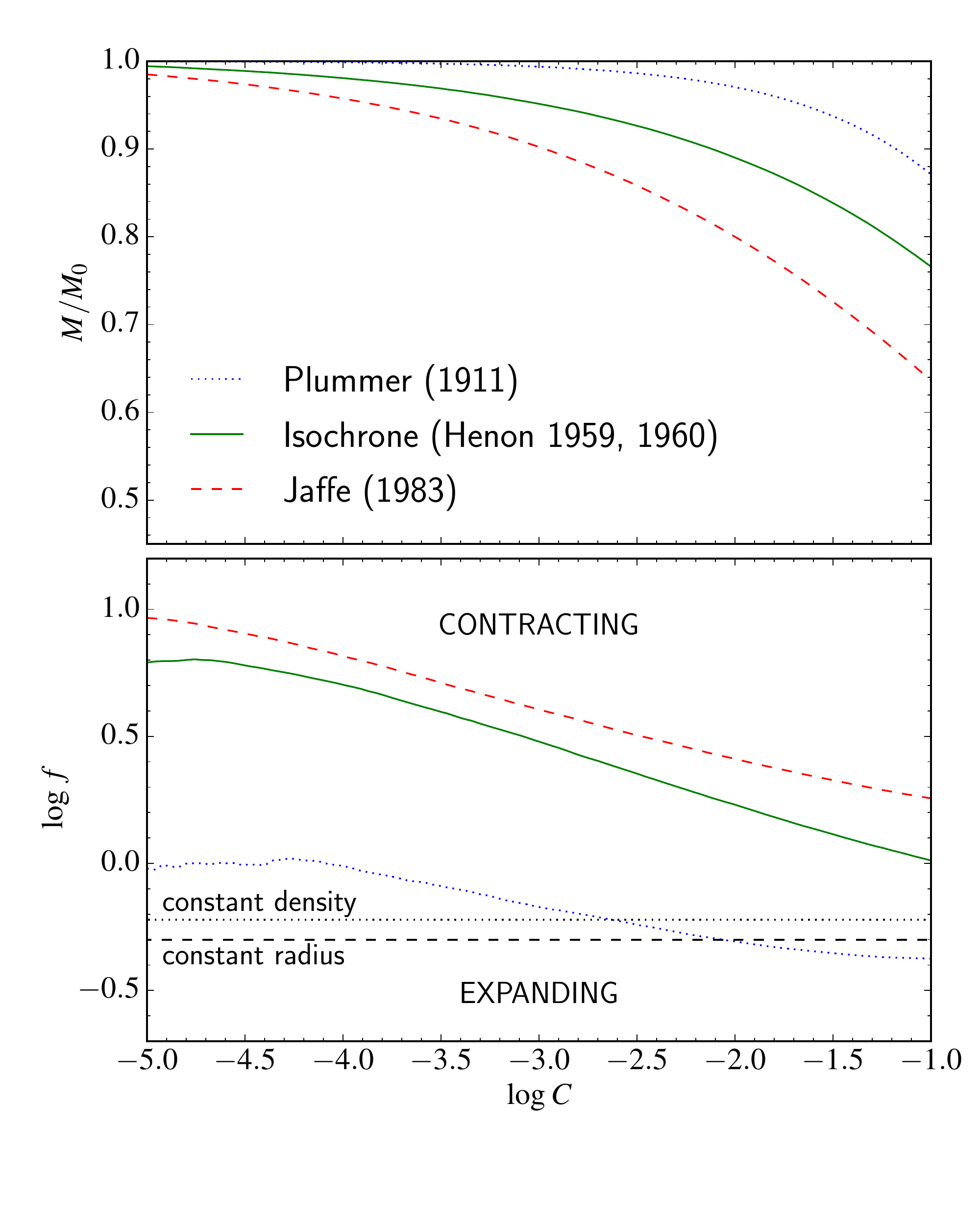}
\caption{Remaining bound mass fraction (top) and logarithmic slope $f$
  of the relation between mass and energy (equation~\ref{eq:dmde})
  after a single tidal shocks $\Delta \E/\langle \phi \rangle =
  C(r/\rh)^2$, where $\langle\phi\rangle$ is the average specific
  potential of stars. Results are computed for three different
  equilibrium models. For the models with density fall-off
  $\rho\propto r^{-4}$ in the outer parts (Jaffe and isochrone models)
  the values of $f$ are similar, and larger than the critical value
  for $f$ to keep $\rhoh$ constant ($f = 3/5$) or $\rh$ constant ($f =
  1/2$). This means that clusters contract to a higher density as the
  result of a tidal shocks.  }
\label{fig:f}
\end{figure}

We add $\Delta \E$ to $\E$ (see Fig.~\ref{fig:schematic_energy}) of
stars in self-consistent, isotropic, equilibrium models and then find
the total $M$ and $E$ of the stars that remain bound (i.e. those for
which $\E+\Delta \E \le 0$). In Fig.~\ref{fig:f} we show the results
for different shock strengths, for the isochrone model
\citep{1959AnAp...22..126H, 1960AnAp...23..474H}, the Jaffe model
\citep{1983MNRAS.202..995J} and the Plummer model
\citep{1911MNRAS..71..460P}, all truncated at $100\rh$.   
\citet*{1987ApJ...323...54E} and
\citet{2003MNRAS.338...85M} find in a sample of young clusters in the
Large Magellanic Cloud that most clusters have luminosity
profiles with logarithmic slopes between $-2.5$ and $-3$ in the outer
regions, corresponding to $-3.5$ and $-4$ after de-projecting under
the assumption of spherical symmetry. The density profiles of the
isochrone and Jaffe models have slopes of $-4$ at large radii, hence
the results for these models are more applicable than that of the
Plummer model which has a steeper density profile ($r^{-5}$).  The
value of $f$ depends on $C$ and hence on $M/M_0$. G06 show that most
of the energy gain is due to encounters with a relative velocity
comparable to the dispersion of the relative velocity distribution,
and with an impact parameter similar to the radius of the GMC (see
their fig.~11). With the results of \citet{1958ApJ...127...17S} and
G06 we find that for such encounters clusters lose a few percent of
their mass, or less. Based on this we adopt $f=3$ from hereon (see
Fig.~\ref{fig:f}). In Section~\ref{ssec:time-scales} we show that the
results are insensitive to the exact value of $f$.

We note that the effect of shocks is strongly self-limiting: if the
mass reduces by a factorof  $q<1$, the density increases by a factor of
$q^{-4}$ (for $f=3$), making the cluster more susceptible against the
next shock.  \citet{1999ApJ...522..935G} discuss this self-limiting
nature of tidal shocks in the context of disc crossings of globular
clusters.  Most studies on GMC interactions implicitly assume that
shocks do not affect the density
\citep[i.e. $f=3/5$,][]{2009ApJ...704..453F}, or only mildly affect
the radius (i.e. $f\simeq1/2$, G06;
\citealt{2010ApJ...712L.184E,2015MNRAS.454.1658K}). In what follows we
show that it is important to include the self-limiting nature of tidal
shocks to understand cluster evolution.

\subsection{time-scale}
\label{ssec:tsh}
We introduce a time-scale $\tshock$ for the change in energy as the
result of repeated shocks

\begin{align}
\frac{\des}{E} &= -\frac{\dr t}{\tshock},
\label{eq:edotshock}
\end{align}
where $\tshock\propto \rhoh$ \citep*[e.g.][]{1958ApJ...127...17S,
  1972ApJ...176L..51O}, such that we can write
  \begin{equation} \tshock = \ggmc
  \frac{\rhoh}{10^2\,\msun\,\pc^{-3}}.
\label{eq:tgmc}
\end{equation}
Here $\ggmc$ is a constant that depends on the dispersion of the
relative velocities between GMCs and the cluster ($\sigrel$), the
surface density of individual GMCs ($\Siggmc$) and the average mass
density of clumpy gaseous structures in the interstellar medium (ISM),
$\rhogmc$, as (\citealt{1958ApJ...127...17S}, G06)
\begin{align}
\ggmc&\simeq 6.5\,\gyr\frac{\sigrel}{10\,\kms}\frac{10\,\msun^2\,\pc^{-5}}{\Siggmc\rhogmc}.
\label{eq:gamma}
\end{align}
There are several constants that have to be chosen to arrive at the
constant of proportionality in equation~(\ref{eq:gamma}) and we used
the values adopted in Section 5.2 of G06\ appropriate for
\citet{1966AJ.....71...64K} models with dimensionless central
potential $W_0 = 7$.  For the Galactic disc ($\sigrel\simeq10\,\kms$,
$\Siggmc\simeq170\,\msun\,\pc^{-2},\rhogmc\simeq0.03\,\msun\,\pc^{-3}$),
we find $\ggmc\simeq 12.8\,\gyr$. Note that the time-scale derived in
G06 was derived for the evolution of the mass, not the energy, for
reasons discussed above. We therefore use their mass-loss time-scale
with the above parameters, and multiply it by $f=3$ (as determined in
Section~\ref{ssec:rhoh_E_shocks}) to get $\ggmc$. This ensures that
clusters lose the same amount of mass as in G06 with our definition of
$\tshock$ for the evolution of $E$.

The increase in density and the decrease in mass as the result of GMC
encounters leads to a reduction of the half-mass relaxation time-scale
($\trh$, see equation~\ref{eq:trh}), hence GMC encounters eventually
push collisionless clusters into the collisional regime.  In the next
section we discuss the effect of two-body relaxation.

\section{Two-body relaxation}
\label{sec:relax}

\subsection{Density evolution}
To describe the effect of two-body relaxation on cluster evolution, we
resort to the model of the evolution of an isolated globular cluster
of \citet[][hereafter H65]{H65}.  Most clusters are confined by a tidal
field, but the model of the isolated cluster describes the early
evolution of clusters that are initially dense compared to their tidal
density \citep*{2011MNRAS.413.2509G}, an assumption we adopt here.

The isolated cluster expands as the result of two-body relaxation with
a central energy source, without losing mass.  Although H\'{e}non's
model is highly idealized, more realistic models that include a
stellar mass spectrum, the mass loss of stars
\citep{2010MNRAS.408L..16G} and stellar-mass black holes
\citep{2013MNRAS.436..584B}, follow similar evolutionary tracks.
Isolated clusters do lose some stars \citep{2002MNRAS.336.1069B}, but
we proceed with the simplifying assumption that the mass remains
constant. The evolution of $\rhoh(E)$ is then simply
(equation~\ref{eq:de})
\begin{equation}
\frac{\dr \rhoh}{\rhoh} = 3\frac{\der}{E}.
\label{eq:Erhoh}
\end{equation}
We note that this result is equivalent to that for tidal shocks for
$f=0$ (equation~\ref{eq:drde_sh}, i.e. shocks that only change the
energy, not the mass).  Shocks affect $M$ more than $E$, such that
$\rhoh$ increases, while $\rhoh$ decreases for two-body relaxation
(equation~\ref{eq:Erhoh}).

\subsection{time-scale}
\label{ssec:trh}
In a relaxation dominated system with a central energy source, the
fractional energy change per $\trh$ is approximately constant
(\citealt{H61}; H65)

\begin{align}
\frac{\der}{E} &= -\zeta\frac{\dr t}{\trh},
\label{eq:edotrelax}
\end{align}
where $\zeta\simeq 0.08-0.10$ for equal-mass clusters (\citealt{H61};
H65; \citealt{2011MNRAS.413.2509G, 2012MNRAS.422.3415A}). For a
constant Coulomb logarithm of $\ln\Lambda = 10$ and a constant stellar
mean mass of $0.5\,\msun$, $\trh$ depends on $M$ and $\rhoh$ as
\begin{align}
\trh = \trs \frac{M}{10^4\,\msun}\left(\frac{\rhoh}{10^2\,\msun\,\pc^{-3}}\right)^{-1/2}.
\label{eq:trh}
\end{align}
For equal-mass systems, $\trs\simeq142\,\myr$
\citep{1971ApJ...164..399S}. A stellar mass spectrum speeds the
relaxation process up by about a factor of 2 for a globular
cluster-like mass function \citep[e.g.][]{1998ApJ...495..786K}. Young
clusters contain more massive stars, causing the two-body relaxation
process to be faster by a factor of 3 (at $\sim100\,\myr$) to 20 (at
$\sim10\,\myr$) than in old globular clusters
\citep{2010MNRAS.408L..16G}.  This effect could be included by making
$\trs$ time dependent, but for consistency with other works we adopt
the value of $\trs$ for equal-mass systems and we adopt a larger
$\zeta=0.5$.

In a collisional system that undergoes tidal shocks, the expansion by
relaxation competes with the shrinking due to shocks and an
equilibrium can be found by considering the respective time-scales of
evolution. This is what we discuss in the next section.

\section{Combined effect of shocks and relaxation}
\label{sec:both}

\subsection{Evolution of the density: an equilibrium MRR}
The density of a cluster evolving under the influence of tidal shocks
and two-body relaxation can be found by adding the change in $E$ due
to tidal shocks (equation~\ref{eq:edotshock}) and relaxation
(equation~\ref{eq:edotrelax})
\begin{align}
\frac{\dr{E}}{E}&= \frac{\des}{E} + \frac{\der}{E}.
\end{align}
We use this in the general expression for the energy
(equation~\ref{eq:de}) and express $\der=(\zeta\tshock/\trh)\des$
(equations~\ref{eq:edotshock} and \ref{eq:edotrelax}). Combined with
the mass evolution due to shocks (equation~\ref{eq:dmde}) we find
\begin{align}
\frac{\dr{\rhoh}}{\rhoh}&=\left(\frac{3}{f}-5 +\frac{3}{f}\frac{\zeta\tshock}{\trh}\right)\frac{\dr{M}}{M}.
\label{eq:drdm_tot}
\end{align}
This relation is equivalent to what we found earlier for shocks
(equations~\ref{eq:dmde} and \ref{eq:drde_sh}), with the additional
contribution of relaxation.

From equation~(\ref{eq:drdm_tot}) we see that for clusters with a low
ratio $\tshock/\trh$ the density increases quickly when $M$ reduces, while
for clusters with a high ratio $\tshock/\trh$ the density decreases
when $M$ decreases.  Because $\tshock/\trh \propto \rhoh^{3/2}/M$,
another way of describing this behaviour is that mass loss (due to shocks) dominates the evolution of low-density, massive clusters, while expansion (due to relaxation) dominates the evolution of dense, low-mass clusters.

The differential equation can be solved via a variable substitution
$\rhoh/M^{2/3}$ and the solution is
\begin{align}
\rhoh &=\left[\frac{AM}{ 1 +  \left(A{\Mi}/\rhohi^{3/2} - 1\right)\left(M/\Mi\right)^{\frac{17}{2}-\frac{9}{2f}}}\right]^{2/3},
\label{eq:rhomsol}
\end{align}
where $\Mi$ and $\rhohi$ are the initial $M$ and $\rhoh$,
respectively, and $A
=0.1\,\msun^{1/2}\pc^{-9/2}\left(17f/9-1\right)\trs/(\zeta\ggmc)$.
This solution is only valid for $f>9/17\simeq0.53$, which includes the
value $f=3$ that we derived in Section~\ref{ssec:rhoh_E_shocks}.

The differential equation given by equation~(\ref{eq:drdm_tot}) has an
attractor solution with $\tshock/\trh
=\,$constant\footnote{\citet{1958AJ.....63..465K} points out that an
  equilibrium radius must exist when cloud interactions and relaxation
  are both at work. However, King assumed that the cluster shrinks as
  the result of stellar ejections, and expands as the result of GMC
  encounters, such that the equilibrium radius is an unstable
  equilibrium and clusters tend to move away from it. Our solution is
  an attractor and clusters will always move towards it. }, i.e. then
$\rhoh\propto M^{2/3}$, and $\rh\propto M^{1/9}$.  Filling in the
parameters for the Milky Way (see Section~\ref{ssec:rhoh_E_shocks}),
with $f=3$ and $\zeta=0.5$, the equilibrium MRR is
\begin{align}
\rh &\simeq 3.8\,\pc\, \left(\frac{\ggmc}{12.8\,\gyr}\right)^{\!\!2/9}\left(\frac{M}{10^4\,\msun}\right)^{1/9},
\label{eq:rh}
 \end{align}
very similar to what is found for young extragalactic cluster
populations in spiral galaxies: $\rh\simeq
3.75\,\pc\,\left(M/10^4\,\msun\right)^{0.1}$ \citep[][where we assumed
  $\rh=(4/3)\reff$ to correct for
  projection]{2004A&A...416..537L}. These clusters have ages
($\lesssim 100\,\myr$) comparable to $\trh$ and $\tshock$, suggesting
that for these clusters, both two-body relaxation and GMC encounters
are important.  G06 and \citet{2010ApJ...712L.184E} use the observed
MRR and density dependent $\tshock$ to argue that cluster lifetimes
depend on mass.  Here we show that such an MRR is in fact the
  result of GMC encounters and two-body relaxation combined for
  clusters with $\trh\simeq \tshock \lesssim \,$Age.

\begin{figure}
\includegraphics[width=8cm]{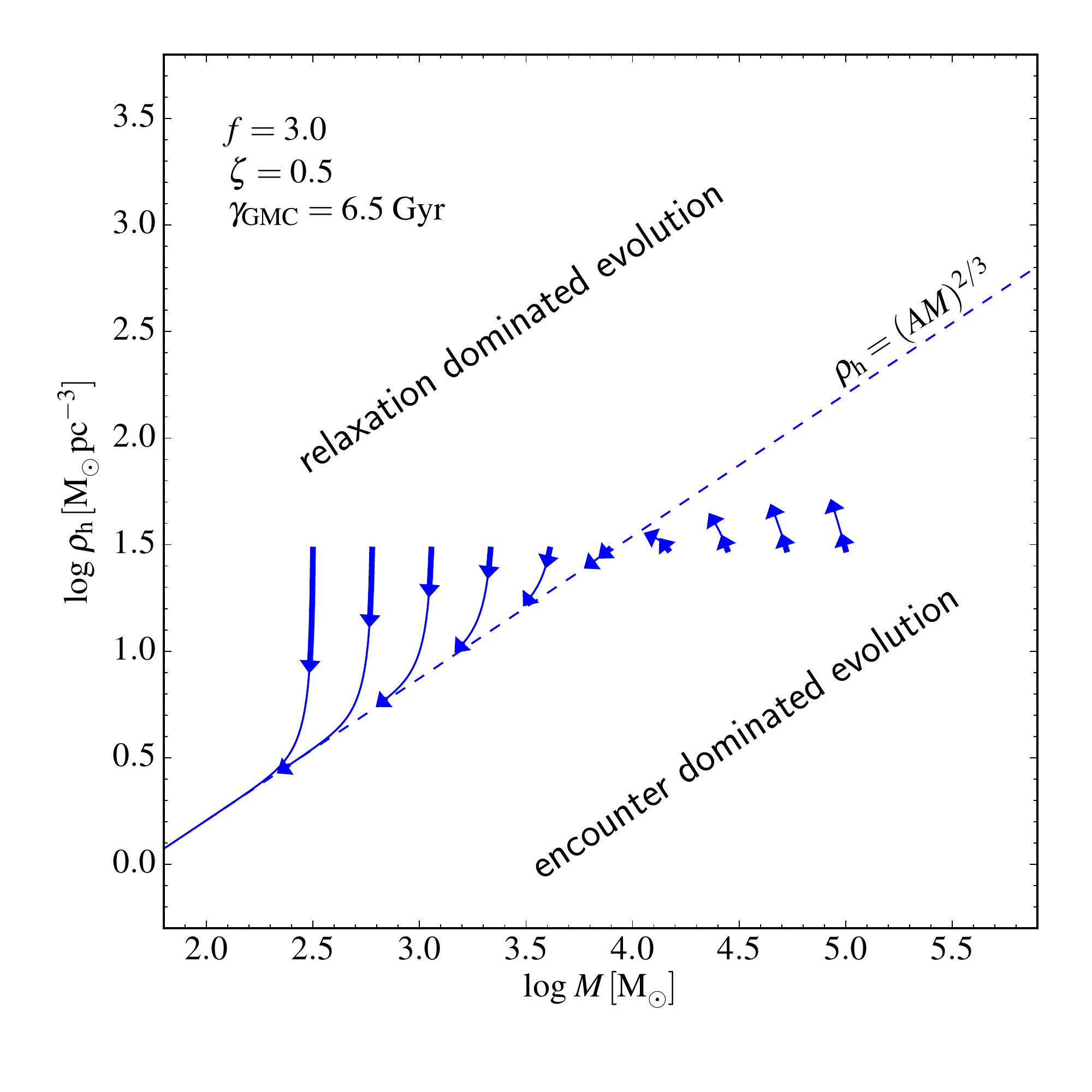}
\caption{Evolution of $\rhoh$ for different initial masses $\Mi$, all
  with the same initial density of $\rhohi=30\,\msun\,\pc^{-3}$.  The
  first arrow along the tracks indicates an age of 30\,\myr, and the
  second arrow marks $300\,\myr$.  For this $\rhohi$, clusters less
  massive than $\log\,\Mi/\msun\simeq4$ are initially denser than the
  equilibrium relation $\rhoh=(AM)^{2/3}$ and expand by two-body
  relaxation towards it. More massive clusters first contract until
  they reach the same relation. }
\label{fig:rhom_schematic}
\end{figure}

In Fig.~\ref{fig:rhom_schematic} we show the evolution of $\rhoh(M)$
for different values of $\Mi$. For all clusters we used
$\rhohi=30\,\msun\,\pc^{-3}$ and with the parameters chosen, the
constant $A\simeq0.02\,\msun^{1/2}\pc^{-9/2}$. The time evolution was
found numerically. From equation~(\ref{eq:rhomsol}) we see that for
this $\rhohi$ clusters with $\Mi\simeq \rhohi^{3/2}/A \simeq
10^4\,\msun$ form on the equilibrium mass-density relation.  For MRRs
steeper than the equilibrium relation, low-mass clusters form in the
relaxation dominated regime of the diagram and expand towards the
equilibrium relation, while more massive clusters form in the shock
dominated regime and contract initially.

\subsection{Dependence of the time-scales on gas properties}
\label{ssec:time-scales}
From the equilibrium MRR we can also derive the relation between
$\tshock$ and $\trh$ and their individual scaling relations
 \begin{align}
  \trh &\simeq \frac{\zeta}{17f/9-1}\tshock\\
\trh &\simeq 302\,\myr\,\left(\frac{\ggmc}{12.8\,\gyr}\right)^{\!\!1/3}\left(\frac{M}{10^4\,\msun}\right)^{2/3}\\
\tshock &\simeq 2822\,\myr\,\left(\frac{\ggmc}{12.8\,\gyr}\right)^{\!\!1/3}\!\left(\frac{M}{10^4\,\msun}\right)^{\!2/3}\label{eq:tgmc2}\\
\tdis &\simeq 940\,\myr\,\left(\frac{\ggmc}{12.8\,\gyr}\right)^{\!\!1/3}\!\left(\frac{M}{10^4\,\msun}\right)^{\!2/3}\label{eq:tdis}.
  \end{align}
Here $\tdis = \tshock/f$ is the time-scale for the evolution of $M$.
{ The dependence of all time-scales on $\ggmc^{1/3}$ shows that
  variations in the properties of molecular gas only mildly affect the
  evolution. } This is because for lower $\ggmc$, the clusters are
denser and lose less mass than what is expected from the linear
dependence of $\tshock$ on $\ggmc$ in equation~(\ref{eq:edotshock}).
The constants of proportionality in equations~(\ref{eq:rh}) and
(\ref{eq:tdis}) depend mildly on the adopted value for $f$, for $f =
[1,3,10]$ they are $[4.3, 3.8, 3.7]\,\pc$ and $[648, 941,
  1032]\,\myr$, respectively. We note that only for $f<9/17$ the
scaling between $\rhoh(M)$ becomes $f$-dependent, because then $\rhoh
\propto M^{\frac{3}{f}-5}$. This excludes the evolution at a constant
density ($f=3/5$), but could allow for evolution at constant
$\rh$ ($f=1/2$).
 
We note that on the equilibrium MRR the dimensionless mass-loss rate
$\xi \equiv -\dot{M}\trh/M = \trh/\tdis \simeq 0.3$ is higher than that
of clusters in isolation \citep[$\xi\simeq0.01$,
  e.g.][]{1998MNRAS.297..794A} or in a static galactic tidal field
\citep[$\xi\simeq0.05$, e.g.][]{H61}.
 
\section{Conclusions}
\label{sec:conclusions}
In this work we show that the interplay between two-body relaxation
and tidal shocks leads to an MRR for star clusters in which the radii
are almost independent of their masses ($\rh\propto M^{1/9}$). Based
on the ISM properties in the solar neighbourhood we estimate that
these processes together could be responsible for the typical cluster
radius of $\sim3\,\pc$ for low-mass ($M\lesssim10^5\,\msun$), young
($\lesssim{\rm few}\,100\,\myr$) star clusters.

The mild dependence of $\rh$ on $M$ implies that the time-scale for
disruption by GMC encounters depends on $M$ as $\tdis\propto M^{2/3}$
(equation~\ref{eq:tdis}).  This mass dependence is similar to what is
found for $\tdis$ as the result of evaporation in a tidal field
($\tdis \propto M^{3/4}$, for a constant Coulomb logarithm and a
constant stellar mass, \citealt{2001MNRAS.325.1323B}) and what was
derived empirically by \citet{2005A&A...441..117L}. This means that
GMC encounters and relaxation contribute in a similar way to `turning
over' a power-law GCMF, as the longer term evaporation process in a
galactic tidal field.

\citet{2010ApJ...712L.184E} uses the high surface densities of
molecular gas in $z\sim2-3$ galaxies
\citep[e.g.][]{2010Natur.463..781T} to argue that $\tdis$ due to cloud
encounters is short enough to turn over the GCMF in the early
evolution of globular clusters. This would imply that all globular
clusters could have formed with a similar power-law mass function as
YMCs in the nearby Universe \citep[power-law
  with index $-2$, see e.g.][for the case of the clusters in the
  Antennae galaxies]{1999ApJ...527L..81Z}.
\citet{2010ApJ...712L.184E} estimates that the value of $\ggmc$ in
equation~(\ref{eq:tgmc}) needs to be a factor of $\sim16$ smaller than
in the solar neighbourhood for this to work in about 500\,\myr. Here
we show that accounting for internal evolution of clusters during this
disruption phase requires $\ggmc$ to be a factor of $16^3=4\,096$
lower instead. Although the observed properties of some high-redshift
galaxies may well be consistent with such short values of $\tdis$, we
note these galaxies are not Milky Way progenitors because they are
more massive than the Milky Way today.

Cluster mass loss as the result of two-body relaxation in the Galactic
tidal field is not sufficient to explain the absence of low-mass GCs
in the outer halo \citep{1998A&A...330..480B, 2001MNRAS.322..247V}.
\citet{2010ApJ...712L.184E} proposes that the additional mass loss
due to tidal shocks with passing GMCs can alleviate this `GCMF
problem'. We demonstrate that $\tdis$ due to GMC encounters has indeed
the correct mass dependence, but we also show that due to the
self-limiting nature of tidal shocks, the required ISM properties for
this to work are more extreme. In a follow-up study we consider this
in more detail, aided by results from hydrodynamical simulations of
Milky Way formation in the cosmological context.

\section*{Acknowledgements}
MG acknowledges support from the Royal Society (University Research
Fellowship) and MG and FR thank the European Research Council
(Starting Grant, grant agreement no. 335936) for support. We
thank Oscar Agertz for interesting discussions and the referees for
comments and suggestions.

\bibliographystyle{mn2e}

\begin{thebibliography}{}

\bibitem[\protect\citeauthoryear{{Aarseth} \& {Heggie}}{{Aarseth} \&
  {Heggie}}{1998}]{1998MNRAS.297..794A}
{Aarseth} S.~J.,  {Heggie} D.~C.,  1998, \mnras, 297, 794

\bibitem[\protect\citeauthoryear{{Alexander} \& {Gieles}}{{Alexander} \&
  {Gieles}}{2012}]{2012MNRAS.422.3415A}
{Alexander} P.~E.~R.,  {Gieles} M.,  2012, \mnras, 422, 3415

\bibitem[\protect\citeauthoryear{{Baumgardt}}{{Baumgardt}}{1998}]{1998A&A...330..480B}
{Baumgardt} H.,  1998, \aap, 330, 480

\bibitem[\protect\citeauthoryear{{Baumgardt}}{{Baumgardt}}{2001}]{2001MNRAS.325.1323B}
{Baumgardt} H.,  2001, \mnras, 325, 1323

\bibitem[\protect\citeauthoryear{{Baumgardt}, {Hut} \& {Heggie}}{{Baumgardt}
  et~al.}{2002}]{2002MNRAS.336.1069B}
{Baumgardt} H.,  {Hut} P.,    {Heggie} D.~C.,  2002, \mnras, 336, 1069

\bibitem[\protect\citeauthoryear{{Binney} \& {Tremaine}}{{Binney} \&
  {Tremaine}}{1987}]{BT1987}
{Binney} J.,  {Tremaine} S.,  1987, {Galactic dynamics}.
Princeton Univ. Press, Princeton, NJ

\bibitem[\protect\citeauthoryear{{Breen} \& {Heggie}}{{Breen} \&
  {Heggie}}{2013}]{2013MNRAS.436..584B}
{Breen} P.~G.,  {Heggie} D.~C.,  2013, \mnras, 436, 584

\bibitem[\protect\citeauthoryear{{Elmegreen}}{{Elmegreen}}{2010}]{2010ApJ...712L.184E}
{Elmegreen} B.~G.,  2010, \apjl, 712, L184

\bibitem[\protect\citeauthoryear{{Elmegreen} \& {Hunter}}{{Elmegreen} \&
  {Hunter}}{2010}]{2010ApJ...712..604E}
{Elmegreen} B.~G.,  {Hunter} D.~A.,  2010, \apj, 712, 604

\bibitem[\protect\citeauthoryear{{Elson}, {Fall} \& {Freeman}}{{Elson}
  et~al.}{1987}]{1987ApJ...323...54E}
{Elson} R.~A.~W.,  {Fall} S.~M.,    {Freeman} K.~C.,  1987, \apj, 323, 54

\bibitem[\protect\citeauthoryear{{Fall}, {Chandar} \& {Whitmore}}{{Fall}
  et~al.}{2009}]{2009ApJ...704..453F}
{Fall} S.~M.,  {Chandar} R.,    {Whitmore} B.~C.,  2009, \apj, 704, 453

\bibitem[\protect\citeauthoryear{{Gieles}, {Baumgardt}, {Heggie} \&
  {Lamers}}{{Gieles} et~al.}{2010}]{2010MNRAS.408L..16G}
{Gieles} M.,  {Baumgardt} H.,  {Heggie} D.~C.,    {Lamers} H.~J.~G.~L.~M.,
  2010, \mnras, 408, L16

\bibitem[\protect\citeauthoryear{{Gieles}, {Heggie} \& {Zhao}}{{Gieles}
  et~al.}{2011}]{2011MNRAS.413.2509G}
{Gieles} M.,  {Heggie} D.~C.,    {Zhao} H.,  2011, \mnras, 413, 2509

\bibitem[\protect\citeauthoryear{{Gieles}, {Portegies Zwart}, {Baumgardt},
  {Athanassoula}, {Lamers}, {Sipior} \& {Leenaarts}}{{Gieles}
  et~al.}{2006}]{2006MNRAS.371..793G}
{Gieles} M.,  {Portegies Zwart} S.~F.,  {Baumgardt} H.,  {Athanassoula} E.,
  {Lamers} H.~J.~G.~L.~M.,  {Sipior} M.,    {Leenaarts} J.,  2006, \mnras, 371,
  793 (G06)

\bibitem[\protect\citeauthoryear{{Gnedin}, {Lee} \& {Ostriker}}{{Gnedin}
  et~al.}{1999}]{1999ApJ...522..935G}
{Gnedin} O.~Y.,  {Lee} H.~M.,    {Ostriker} J.~P.,  1999, \apj, 522, 935

\bibitem[\protect\citeauthoryear{{Gnedin} \& {Ostriker}}{{Gnedin} \&
  {Ostriker}}{1999}]{1999ApJ...513..626G}
{Gnedin} O.~Y.,  {Ostriker} J.~P.,  1999, \apj, 513, 626

\bibitem[\protect\citeauthoryear{{H\'{e}non}}{{H\'{e}non}}{1959}]{1959AnAp...22..126H}
{H\'{e}non} M.,  1959, Annales d'Astrophysique, 22, 126

\bibitem[\protect\citeauthoryear{{H{\'e}non}}{{H{\'e}non}}{1960}]{1960AnAp...23..474H}
{H{\'e}non} M.,  1960, Annales d'Astrophysique, 23, 474

\bibitem[\protect\citeauthoryear{{H{\'e}non}}{{H{\'e}non}}{1961}]{H61}
{H{\'e}non} M.,  1961, Annales d'Astrophysique, 24, 369;  translation:
  ArXiv:1103.3499

\bibitem[\protect\citeauthoryear{{H{\'e}non}}{{H{\'e}non}}{1965}]{H65}
{H{\'e}non} M.,  1965, Annales d'Astrophysique, 28, 62; translation:
  ArXiv:1103.3498 (H65)

\bibitem[\protect\citeauthoryear{{Jaffe}}{{Jaffe}}{1983}]{1983MNRAS.202..995J}
{Jaffe} W.,  1983, \mnras, 202, 995

\bibitem[\protect\citeauthoryear{{Kharchenko}, {Piskunov}, {R{\"o}ser},
  {Schilbach} \& {Scholz}}{{Kharchenko} et~al.}{2005}]{2005A&A...438.1163K}
{Kharchenko} N.~V.,  {Piskunov} A.~E.,  {R{\"o}ser} S.,  {Schilbach} E.,
  {Scholz} R.-D.,  2005, \aap, 438, 1163

\bibitem[\protect\citeauthoryear{{Kim}, {Lee} \& {Goodman}}{{Kim}
  et~al.}{1998}]{1998ApJ...495..786K}
{Kim} S.~S.,  {Lee} H.~M.,    {Goodman} J.,  1998, \apj, 495, 786

\bibitem[\protect\citeauthoryear{{King}}{{King}}{1958}]{1958AJ.....63..465K}
{King} I.,  1958, \aj, 63, 465

\bibitem[\protect\citeauthoryear{{King}}{{King}}{1966}]{1966AJ.....71...64K}
{King} I.~R.,  1966, \aj, 71, 64

\bibitem[\protect\citeauthoryear{{Kruijssen}}{{Kruijssen}}{2015}]{2015MNRAS.454.1658K}
{Kruijssen} J.~M.~D.,  2015, \mnras, 454, 1658

\bibitem[\protect\citeauthoryear{{Kundic} \& {Ostriker}}{{Kundic} \&
  {Ostriker}}{1995}]{1995ApJ...438..702K}
{Kundic} T.,  {Ostriker} J.~P.,  1995, \apj, 438, 702

\bibitem[\protect\citeauthoryear{{Lamers}, {Gieles}, {Bastian}, {Baumgardt},
  {Kharchenko} \& {Portegies Zwart}}{{Lamers}
  et~al.}{2005}]{2005A&A...441..117L}
{Lamers} H.~J.~G.~L.~M.,  {Gieles} M.,  {Bastian} N.,  {Baumgardt} H.,
  {Kharchenko} N.~V.,    {Portegies Zwart} S.,  2005, \aap, 441, 117

\bibitem[\protect\citeauthoryear{{Larsen}}{{Larsen}}{2004}]{2004A&A...416..537L}
{Larsen} S.~S.,  2004, \aap, 416, 537

\bibitem[\protect\citeauthoryear{{Larson}}{{Larson}}{1981}]{1981MNRAS.194..809L}
{Larson} R.~B.,  1981, \mnras, 194, 809

\bibitem[\protect\citeauthoryear{{Mackey} \& {Gilmore}}{{Mackey} \&
  {Gilmore}}{2003}]{2003MNRAS.338...85M}
{Mackey} A.~D.,  {Gilmore} G.~F.,  2003, \mnras, 338, 85

\bibitem[\protect\citeauthoryear{{Ostriker}, {Spitzer} \&
  {Chevalier}}{{Ostriker} et~al.}{1972}]{1972ApJ...176L..51O}
{Ostriker} J.~P.,  {Spitzer} L.~J.,    {Chevalier} R.~A.,  1972, \apjl, 176,
  L51

\bibitem[\protect\citeauthoryear{{Plummer}}{{Plummer}}{1911}]{1911MNRAS..71..460P}
{Plummer} H.~C.,  1911, \mnras, 71, 460

\bibitem[\protect\citeauthoryear{{Portegies Zwart}, {McMillan} \&
  {Gieles}}{{Portegies Zwart} et~al.}{2010}]{2010ARA&A..48..431P}
{Portegies Zwart} S.~F.,  {McMillan} S.~L.~W.,    {Gieles} M.,  2010, \araa,
  48, 431

\bibitem[\protect\citeauthoryear{{Scheepmaker}, {Haas}, {Gieles}, {Bastian},
  {Larsen} \& {Lamers}}{{Scheepmaker} et~al.}{2007}]{2007A&A...469..925S}
{Scheepmaker} R.~A.,  {Haas} M.~R.,  {Gieles} M.,  {Bastian} N.,  {Larsen}
  S.~S.,    {Lamers} H.~J.~G.~L.~M.,  2007, \aap, 469, 925

\bibitem[\protect\citeauthoryear{{Spitzer}}{{Spitzer}}{1987}]{1987degc.book.....S}
{Spitzer} L.,  1987, {Dynamical evolution of globular clusters}.
Princeton, NJ, Princeton University Press, 1987, 191 p.

\bibitem[\protect\citeauthoryear{{Spitzer}}{{Spitzer}}{1958}]{1958ApJ...127...17S}
{Spitzer} L.~J.,  1958, \apj, 127, 17

\bibitem[\protect\citeauthoryear{{Spitzer} \& {Hart}}{{Spitzer} \&
  {Hart}}{1971}]{1971ApJ...164..399S}
{Spitzer} L.~J.,  {Hart} M.~H.,  1971, \apj, 164, 399

\bibitem[\protect\citeauthoryear{{Tacconi}, {Genzel}, {Neri}, {Cox}, {Cooper},
  {Shapiro}, {Bolatto} \& {Bouch{\'e}}}{{Tacconi}
  et~al.}{2010}]{2010Natur.463..781T}
{Tacconi} L.~J.  et al.   2010, \nat, 463, 781

\bibitem[\protect\citeauthoryear{{Terlevich}}{{Terlevich}}{1987}]{1987MNRAS.224..193T}
{Terlevich} E.,  1987, \mnras, 224, 193

\bibitem[\protect\citeauthoryear{{Urquhart}, {Moore}, {Csengeri}, {Wyrowski},
  {Schuller}, {Hoare}, {Lumsden}, {Mottram} \& {et al.}}{{Urquhart}
  et~al.}{2014}]{2014MNRAS.443.1555U}
{Urquhart} J.~S.  {et
  al.} 2014, \mnras, 443, 1555

\bibitem[\protect\citeauthoryear{{van den Bergh}}{{van den
  Bergh}}{2006}]{2006AJ....131.1559V}
{van den Bergh} S.,  2006, \aj, 131, 1559

\bibitem[\protect\citeauthoryear{{Vesperini}}{{Vesperini}}{2001}]{2001MNRAS.322..247V}
{Vesperini} E.,  2001, \mnras, 322, 247

\bibitem[\protect\citeauthoryear{{Weinberg}}{{Weinberg}}{1994}]{1994AJ....108.1398W}
{Weinberg} M.~D.,  1994, \aj, 108, 1398

\bibitem[\protect\citeauthoryear{{Wielen}}{{Wielen}}{1985}]{1985IAUS..113..449W}
{Wielen} R.,  1985, in {Goodman} J.,  {Hut} P.,  eds, Proc. IAU Symp. 113, Dynamics
  of Star Clusters {Dynamics of open star clusters}. Reidel, Dordrecht,
p 449

\bibitem[\protect\citeauthoryear{{Zepf}, {Ashman}, {English}, {Freeman} \&
  {Sharples}}{{Zepf} et~al.}{1999}]{1999AJ....118..752Z}
{Zepf} S.~E.,  {Ashman} K.~M.,  {English} J.,  {Freeman} K.~C.,    {Sharples}
  R.~M.,  1999, \aj, 118, 752

\bibitem[\protect\citeauthoryear{{Zhang} \& {Fall}}{{Zhang} \&
  {Fall}}{1999}]{1999ApJ...527L..81Z}
{Zhang} Q.,  {Fall} S.~M.,  1999, \apjl, 527, L81

\end{thebibliography}

\bsp	
\label{lastpage}
\end{document}